# The nanophysics age and its new perspectives

Catalano E[1]

[1]University of Milano Bicocca, Piazza dell'Ateneo Nuovo, 1, 20126 Milan, Italy

The nanophysics is halfway between the size scales of quantum mechanics and macroscopic physics governed by the laws of Newton and Einstein. The correct definition of nanophysics is the physics of structures and artefacts with dimensions in the nanometer range or of phenomena occurring in nanoseconds [1].

Modern physical methods whose fundamental are developed in physics laboratories have become critically important in nanoscience. Nanophysics brings together multiple disciplines, using theoretical and experimental methods to determine the physical properties of materials in the nanoscale size range. Interesting properties include the structural, electronic, optical, and thermal behavior of nanomaterials; electrical and thermal conductivity; the forces between nanoscale objects; and the transition between classical and quantum behavior. Nanophysics has now become an independent branch of physics, simultaneously expanding into many new areas and playing a vital role in fields that were once the domain of engineering, chemical, or life sciences [1].

Nanoscience and nanotechnology are all about relating and exploiting phenomena for materials having one, two or three dimensions reduced to the nanoscale. Breakthroughs in nanotechnology require a firm grounding in the principles of nanophysics. It is intended to fulfill a crucial purpose. Nanophysics aims to connect scientists with disparate interests to begin interdisciplinary projects and incorporate the theory and methodology of other fields into their work [2].

Their evolution may be related to three exciting happenings that took place in a short span from the early to mid-1980s with the award of Nobel prizes to each of them [2]. These were: (i) the discovery quantum Hall effect in a two-dimensional electron gas; (ii) the invention of scanning tunnelling microscopy (STM); and (iii) the discovery of fullerene as the new form of carbon. The latter two, within a few years, further led to the remarkable invention of the atomic force microscope (AFM) and, in the early 1990s the extraordinary discovery of carbon nanotubes (CNT), which soon provided the launch pad for the present-day nanotechnology [2]. The STM and AFM have emerged as the most powerful tools to examine, control and manipulate matter at the atomic, molecular and macromolecular scales and these functionalities constitute the mainstay of nanotechnology. Interestingly, this exciting possibility of nanolevel tailoring of materials was envisioned way back in 1959 by Richard Feynman in his lecture, "There's plenty of room at the bottom" [3].

**Nanophysics applications**

When things get small or cold (or both!), quantum effects start to appear. Nanophysics develops various devices and instruments to reveal and quantify them.

Novel materials, structures and devices are constructed through a variety of fabrication techniques, including e-beam lithography, focused-ion-beam milling, nano-manipulation, and self-assembly [1]. They are then tested at temperatures ranging from ambient down to a few tens of millikelvin using various probes, microscopes and cryostats. Probing the form and function of nano-structure and devices requires and inspires the development of ultra-sensitive detectors, sources (of quanta) and microscopes [1].

**Quantum Measurements using Nanomechanical Resonators**

The electron has dominated technology, measurement, communications and information processing for around one century. Hard limits may restrict its future dominance. One promising disruptive technology that may grow in future is based on NEMS (nano-scale electromechanical system).

Resonators based on NEMS (so called NMRs) are expected to have a range of applications, from ultra-sensitive sensors for mass, force, charge, spin and chemical specificity, through single-molecule bio-sensing, information storage and processing technologies, to nanoscale refrigerators.

They are sufficiently small that mesoscopic quantum mechanical behaviour is expected to appear, at low temperatures or even at room temperature, with all of the quantum metrology capabilities that have hither to been found in atomic and condensed matter physics.

There is a key requirement to extend quantum metrology to the nanoscale and to achieve measurements that are limited only by counting statistics or, going further, by the Heisenberg uncertainty principle limit.

It focuses on metrological aspects of NMRs as they approach the quantum regime, where the resonator's state is not significantly 'mixed' by thermal noise; in other words, one requires

$$\hbar\omega > k_B T$$

where $\omega$ is one of the resonator's fundamental frequencies, $T$ is its effective operating temperature, and $\hbar$ and $k_B$ are fundamental constants.

The ultimate metrological target, towards which it is aiming, is the provision of robust, convenient quantum-NEMS-based devices for the generation and counting of individual phonons. If this can be achieved it will have the mechanical analogue of 'quantum optics' and a whole new area of device engineering.

To avoid the need for ultra-low temperatures an NMR should operate at the highest possible frequency. It should also be capable of fabrication on the smallest possible length scale. A number of nanofabricated structures were researched composed of single-crystal materials such as Si, $Si_3N_4$ and particularly carbon nanotubes [4-7].

**Nanomaterials and Nanomedicine**

Various applications of nanoscale science to the field of medicine have resulted in the ongoing development of the subfield of nanomedicine. Nanomedicine is a relatively new field that is rapidly evolving. Formulation of drugs on the nanoscale imparts many physical and biological advantages. Such advantages can in turn translate into improved therapeutic efficacy and reduced toxicity [8].

Additionally, there is now broad consensus among medical researchers and practitioners that along with personalized medicine and regenerative medicine, nanomedicine is likely to revolutionize our definitions of what constitutes human disease and its treatment. Nanomaterials are defined as the production of matter with at least one dimension ranging between 1 and 100 nanometers. Due to the very small size and the resulting high surface/volume ratio, nanomaterials have physical-chemical properties that differ from those of macroscopic materials. Nowdays nanomaterials are often applied in many industrial fields including electronics, optics, textile and many others till to biomedicine [8].

The use of nanoparticles (NPs) in medicine has expanded recently, especially in diagnostic [9]. Actually nanoparticles could be designed as contrast agents, targeted therapies in cancer or nanocarriers able to bind, specifically transport biomolecules and accumulate to the site to treat.